\documentclass[letterpaper,journal]{IEEEtran}

\usepackage[utf8]{inputenc}
\usepackage[T1]{fontenc}
\usepackage{amsthm,amsmath,amssymb,amsfonts,mathtools}
\usepackage{algorithmic}
\usepackage{algorithm}
\usepackage{array}
\usepackage[caption=false,font=normalsize,labelfont=sf,textfont=sf]{subfig}
\usepackage{textcomp}
\usepackage{stfloats}
\usepackage{url}
\usepackage{verbatim}
\usepackage{graphicx}
\usepackage{cite}
\usepackage{xcolor}
\usepackage{booktabs}
\usepackage{multirow}

\newcommand{\C}{\mathbb{C}}

\newcommand{\Id}{\mathbf{I}}
\newcommand{\vect}{\operatorname{vec}}

\newcommand{\argmin}{\operatorname*{arg\,min}}

\newcommand{\snrdb}{\mathrm{SNR}_{\mathrm{dB}}}
\newcommand{\nmse}{\mathrm{NMSE}}
\newcommand{\rhoGrid}{\mathcal{G}_{\rho}}
\newcommand{\cn}{\mathcal{CN}}

\begin{document}
	
	\title{Adaptive Plug-and-Play Channel Estimation with Consistency Models for MIMO Systems}
	
	\author{Jinlong Li, Kexin Fang,~\IEEEmembership{Graduate Student Member,~IEEE}, Peng Yang,~\IEEEmembership{Member,~IEEE}, 
		Zehui Xiong,~\IEEEmembership{Senior Member,~IEEE}, 
		Xianbin Cao,~\IEEEmembership{Senior Member,~IEEE}, 
		and Dapeng Oliver Wu,~\IEEEmembership{Fellow,~IEEE}

		\thanks{J. Li, P. Yang and X. Cao are with School of Electronic and Information Engineering, Beihang University, Beijing 100191, China. (e-mail: \{li\_j\_l;peng\_yang;xbcao\}@buaa.edu.cn)
			
			Z. Xiong is with the School of Electronics, Electrical Engineering and Computer Science, Queen's University Belfast, Belfast BT7 1NN, United Kingdom. (e-mail: z.xiong@qub.ac.uk)
			
			K. Fang, and D. O. Wu are with the Department of Computer Science, City University of Hong Kong, Kowloon, Hong Kong, China. (e-mail: fangkexinwonder@163.com, dpwu@ieee.org)
		}
	}
	\maketitle
	
	\begin{abstract}
		
		This paper proposes a consistency-model-based channel estimation algorithm for multiple-input multiple-output (MIMO) systems. The proposed algorithm employs a consistency model (CM) to learn the angle-domain channel distribution and uses the trained CM as a plug-and-play (PnP) generative prior for MIMO channel estimation. The proposed algorithm alternates between a pilot-observation-based data-consistency update and a CM-prior-based denoising update. In addition, the proposed algorithm adaptively selects the penalty parameter according to residual energy and residual whiteness, and adjusts the CM denoising level according to the observed signal-to-noise ratio (SNR), thereby avoiding the performance degradation caused by fixed inference schedules under varying observation conditions. Simulation results show that the proposed algorithm not only reduces the number of inference steps by \(50\%\)--\(90\%\), but also achieves high estimation accuracy and favorable cross-dataset performance.

	\end{abstract}
	

	\section{Introduction}
	
	Generative models can learn complex data distributions, making them natural candidates for learning channel priors in multiple-input multiple-output (MIMO) channel estimation. Compared with handcrafted priors, which are usually tied to prescribed sparsity or parametric assumptions, generative models can learn realistic and complex channel features from representative datasets. Various generative models have been explored for channel estimation, including generative adversarial networks (GANs), Gaussian mixture models (GMMs), and variational autoencoders (VAEs). Among them, diffusion models (DMs) have become particularly attractive because of their stable training behavior and strong generative capability \cite{zhou2026generative,ho2020denoising}.
	
	However, applying DMs in practical channel estimation can be computationally expensive. Channel recovery must proceed along a reverse diffusion or probability-flow trajectory, which leads to multiple neural function evaluations (NFEs). Recent work has attempted to reduce this cost. The algorithm in \cite{fesl2024diffusion} uses a lightweight angular-domain network and signal-to-noise-ratio (SNR)-matched inference to skip reverse steps that are inconsistent with the observation SNR. DiffPace combines a DM prior with plug-and-play (PnP) estimation and ordinary differential equation (ODE) inference for mmWave and terahertz ultra-massive MIMO channel estimation \cite{hu2026diffpace}. Although these designs improve inference efficiency, channel recovery still relies on a multi-step reverse or ODE procedure. It is therefore necessary to explore more efficient generative models for low-latency channel estimation.
	
	Consistency models (CMs) provide a different route to fast generative inference. Unlike DMs that denoise gradually along a reverse trajectory, CMs learn to map any noisy state on a probability-flow trajectory directly to a common clean endpoint, and therefore naturally support one-step or few-step generation \cite{song2023consistency,song2023improved}. This property matches the low-complexity and low-latency requirements of channel estimation, making CMs a promising alternative for constructing fast generative channel priors. However, to the best of our knowledge, CMs have not yet been applied to MIMO channel estimation.
	
	Recent work has also explored CMs for general inverse problems. CM4IR combines CMs with improved initialization, back-projection guidance, and noise injection for zero-shot image restoration \cite{garber2025zero}. However, because this algorithm relies on pseudo-inverse/back-projection operations, it is not straightforward to extend it to nonlinear or severely ill-conditioned observation settings. PnP-CM instead treats CM denoisers as proximal operators of a prior and embeds them into the integrated framework of PnP and alternating direction method of multipliers (ADMM) \cite{gulle2025pnpcm}. Its performance, however, largely depends on empirically selected hyperparameter schedules, and fixed empirical schedules are difficult to adapt to the varying SNRs and pilot configurations encountered in communication receivers.
	
	Therefore, this paper proposes a CM-based channel estimation algorithm with adaptive parameter selection. The proposed algorithm automatically adjusts the penalty parameter and the CM denoising level according to residual statistics and the observed SNR, thereby reducing manual tuning and improving robustness under compressed pilot observations.
	
	The main contributions of this work are as follows:
	\begin{itemize}
		\item We propose a CM-based channel estimation algorithm for MIMO systems. We designed and trained the CM for the algorithm to learn the angle-domain channel distribution, and used the learned model as a PnP generative prior for channel estimation.
		\item We design an adaptive parameter-selection mechanism for the proposed algorithm. The penalty parameter is determined according to residual-energy consistency and residual-whiteness criteria, while the CM denoising level is jointly determined by the SNR-dependent regularization strength and the penalty parameter.
		\item We validate the effectiveness and robustness of the proposed algorithm through simulations. The results show that the proposed algorithm achieves high estimation accuracy with only \(5\) to \(10\) NFEs.
		
	\end{itemize}

	\section{Preliminaries}
	\label{sec:preliminaries}
	
	\subsection{MIMO System Model}
	
	Consider a narrowband MIMO training system where both the transmitter and receiver use uniform linear arrays (ULAs). Let \(N_t\) and \(N_r\) denote the numbers of transmit and receive antennas, respectively. Let \(\mathbf{H}\in\C^{N_r\times N_t}\) denote the spatial-domain channel, \(\mathbf{P}\in\C^{N_t\times M_t}\) the transmit pilot matrix, and \(\mathbf{W}\in\C^{N_r\times M_rL_r}\) the stacked receive combining matrix, where \(L_r\) is the number of radio-frequency (RF)-chain outputs in each receive scan. The received pilot matrix is
	\begin{equation}
		\mathbf{Y}=\mathbf{W}^{H}\mathbf{H}\mathbf{P}+\mathbf{N}.
	\end{equation}
	Here, \(\mathbf{Y}\in\C^{M_rL_r\times M_t}\), and \(\mathbf{N}\) is the effective observation noise. Using unitary discrete Fourier transform (DFT) dictionaries \(\mathbf{F}_t\in\C^{N_t\times N_t}\) and \(\mathbf{F}_r\in\C^{N_r\times N_r}\), the angle-domain channel is defined as
	\begin{equation}
		\mathbf{H}_a=\mathbf{F}_r^H\mathbf{H}\mathbf{F}_t.
	\end{equation}
	Equivalently, \(\mathbf{H}=\mathbf{F}_r\mathbf{H}_a\mathbf{F}_t^H\). Substituting this representation into the received signal and vectorizing the result gives the linear inverse model
	\begin{equation}
		\mathbf{y}=\mathbf{A}\mathbf{h}+\mathbf{n},
		\label{eq:linear-model}
	\end{equation}
	where \(\mathbf{h}=\vect(\mathbf{H}_a)\in\C^{N_tN_r}\), \(\mathbf{y}=\vect(\mathbf{Y})\), \(\mathbf{n}=\vect(\mathbf{N})\), and
	\begin{equation}
		\mathbf{A}
		=
		\left(\mathbf{F}_t^H\mathbf{P}\right)^T
		\otimes
		\left(\mathbf{W}^H\mathbf{F}_r\right).
	\end{equation}

	We assume \(\mathbf{n}\sim\cn(\mathbf{0},\sigma_n^2\Id)\). In practical receivers, \(\sigma_n^2\) and the operating SNR can be obtained from pilot-aided or decision-directed noise/SNR estimation modules \cite{boumard2003novel}, and are treated in this paper as available side information.
	
	Estimating the angle-domain channel from \(\mathbf{y}\) is therefore formulated as a regularized inverse problem. The channel estimate is obtained by minimizing
	\begin{equation}
		\widehat{\mathbf{h}}
		=
		\argmin_{\mathbf{x}}
		\frac{1}{2}\|\mathbf{y}-\mathbf{A}\mathbf{x}\|_2^2+\lambda g(\mathbf{x}),
		\label{eq:reg-problem}
	\end{equation}
	where the squared-error term \(\frac{1}{2}\|\mathbf{y}-\mathbf{A}\mathbf{x}\|_2^2\) is the data-fidelity term that enforces agreement between the estimate and the pilot observation model, \(g(\mathbf{x})\) is a regularization term or prior on the channel, and \(\lambda>0\) balances data consistency and prior strength.
	
	\subsection{Consistency Models}
	
	CMs learn a mapping from an intermediate noisy state on a probability-flow trajectory to a common clean endpoint \cite{song2023consistency,song2023improved}. Unlike diffusion models that usually require multi-step reverse sampling, CMs support one-step or few-step denoising, which makes them suitable as low-latency generative priors. For a channel vector \(\mathbf{h}_0\) from the training distribution and its perturbed version \(\mathbf{h}_t\) at noise level \(t\in[\epsilon,T]\), the CM mapping used in this paper is written as
	\begin{equation}
		f_\theta:\left(\mathbf{h}_t,t\right)\mapsto \mathbf{h}_{\epsilon},
		\label{eq:cm-map}
	\end{equation}
	where \(\epsilon\) is the minimum noise level near the data endpoint.
	
	The model is trained to satisfy the self-consistency property, i.e., states from the same probability-flow trajectory should be mapped to the same endpoint, while the mapping reduces to the identity near the minimum noise level. CMs can be obtained by distilling a pretrained diffusion model or by direct consistency training \cite{song2023improved}.

	\section{Proposed Channel Estimation Algorithm}
	\label{sec:method}
	
	To solve the regularized inverse problem in \eqref{eq:reg-problem}, we learn a channel prior and incorporate it into model-based optimization. Specifically, ADMM decomposes \eqref{eq:reg-problem} into a data-consistency subproblem handled by a linear update and a prior subproblem handled by the CM prior. For a received signal at an arbitrary SNR, the proposed algorithm automatically determines the penalty parameter and the CM denoising level at each iteration, instead of relying on a fixed empirical schedule.
	
	\subsection{ADMM Reformulation and Algorithm Update}
	
	Introducing variables \(\mathbf{z}\) and \(\mathbf{x}\) with the constraint \(\mathbf{z}=\mathbf{x}\), \eqref{eq:reg-problem} can be rewritten as
	\begin{equation}
		\min_{\mathbf{z},\mathbf{x}}
	    \frac{1}{2}\|\mathbf{y}-\mathbf{A}\mathbf{z}\|_2^2+\lambda g(\mathbf{x}),
		\quad \mathrm{s.t.}\quad \mathbf{z}=\mathbf{x}.
		\label{eq:split-problem}
	\end{equation}

	Using the scaled dual variable \(\boldsymbol{\mu}_k\), define \(\widetilde{\mathbf{z}}_k=\mathbf{x}_k-\boldsymbol{\mu}_k\) and \(\widetilde{\mathbf{x}}_k=\mathbf{z}_{k+1}+\boldsymbol{\mu}_k\). The basic ADMM iterations are
	\begin{subequations}
		\label{eq:admm-basic}
		\begin{align}
			\mathbf{z}_{k+1}
			&=
			\argmin_{\mathbf{z}}
			\frac{1}{2}\|\mathbf{y}-\mathbf{A}\mathbf{z}\|_2^2+
			\frac{\rho_k}{2}\|\mathbf{z}-\widetilde{\mathbf{z}}_k\|_2^2, \label{eq:admm-basic-z}\\
			\mathbf{x}_{k+1}
			&=
			\argmin_{\mathbf{x}}
			\lambda g(\mathbf{x})+
			\frac{\rho_k}{2}\|\widetilde{\mathbf{x}}_k-\mathbf{x}\|_2^2, \label{eq:admm-basic-x}\\
			\boldsymbol{\mu}_{k+1}
			&=
			\boldsymbol{\mu}_k+\mathbf{z}_{k+1}-\mathbf{x}_{k+1}.
			\label{eq:admm-basic-mu}
		\end{align}
	\end{subequations}

	In the implementation, momentum is subsequently introduced by replacing \(\mathbf{x}_k\) and \(\boldsymbol{\mu}_k\) in the reference variables with their extrapolated versions.
	
	The \(\mathbf{z}\)-subproblem is quadratic and has the following closed-form solution:
	\begin{equation}
		\mathbf{z}_{k+1}
		=
		\left(\mathbf{A}^H\mathbf{A}+\rho_k\Id\right)^{-1}
		\left(\mathbf{A}^H\mathbf{y}+\rho_k\widetilde{\mathbf{z}}_k\right).
		\label{eq:z-update}
	\end{equation}

	The \(\mathbf{x}\)-subproblem is the proximal operator of the prior:
	\begin{equation}
		\mathbf{x}_{k+1}
		=
		\operatorname{prox}_{\lambda g/\rho_k}
		\left(\widetilde{\mathbf{x}}_k\right),
		\label{eq:x-prox}
	\end{equation}
	and the theoretical proximal denoising scale is \(\sqrt{\lambda/\rho_k}\). In PnP-CM, this proximal step is replaced with a trained CM denoiser \(f_{\theta}\):
	\begin{equation}
		\mathbf{x}_{k+1}
		=
		f_{\theta}\left(\widetilde{\mathbf{x}}_k,t_k\right).
		\label{eq:cm-update}
	\end{equation}
	
	The original PnP-CM algorithm recommends injecting controlled random perturbations before the CM update, and the CM denoising level, penalty parameter, and momentum parameter sequences at each iteration are all predetermined empirically \cite{gulle2025pnpcm}. However, a fixed hyperparameter sequence cannot adapt to varying SNR conditions in channel estimation. Moreover, for each received observation, we expect the channel estimate to be relatively stable. Therefore, the proposed algorithm does not actively inject noise, and selects \(\rho_k\) and \(t_k\) adaptively from the current residual and SNR.
	
	\subsection{Adaptive Inference Details}
	\label{subsec:adaptive-inference}
	
	At each outer iteration, we select the ADMM penalty parameter \(\rho_k\) according to residual statistics. For any candidate \(\rho>0\), the data-consistency update in \eqref{eq:z-update} gives the corresponding estimate \(\mathbf{z}_{k+1}(\rho)\), and its observation residual is defined as
	\begin{equation}
		\mathbf{r}_{k+1}(\rho)
		=
		\mathbf{A}\mathbf{z}_{k+1}(\rho)-\mathbf{y},
		\qquad
		\mathbf{r}_{k+1}(\rho)\in\C^m,
		\label{eq:candidate-residual}
	\end{equation}
	where \(m=M_tM_rL_r\) is the dimension of the observation vector. When \(\rho\) is properly selected, the residual should be dominated mainly by measurement noise. Hence, a statistically reasonable residual should not only have an average energy consistent with the noise variance, but also be as uncorrelated as possible at nonzero lags. The former corresponds to the classical discrepancy principle \cite{morozov1966regularization}, while the latter follows the basic idea of residual-whiteness diagnostics \cite{pragliola2023admm}.
	
	Based on this observation, we adopt a two-stage rule for selecting \(\rho_k\). The first stage enforces residual energy consistency. Define the residual-energy consistency objective \(E(\rho)\) as
	\begin{equation}
		E(\rho)
		=
		\left|
		\frac{\|\mathbf{r}_{k+1}(\rho)\|_2^2/m}
		{\sigma_n^2}
		-1
		\right|.
		\label{eq:energy-objective}
	\end{equation}
	Here, \(E(\rho)\) measures the relative mismatch between the mean squared residual energy and the noise power. However, energy matching only constrains the zero-lag second-order statistic of the residual, and cannot rule out structured mismatch remaining in the residual. In addition, due to finite-sample fluctuations, noise-variance estimation errors, and the use of a discrete candidate grid, forcing the residual energy to be exactly equal to the noise power can make parameter selection overly sensitive and numerically unstable. We therefore define an energy-compatible interval using a preset tolerance \(\eta\), and retain only candidates that are statistically consistent with the noise energy. The energy-consistent set is
	\begin{equation}
		\mathcal{C}_{\eta}
		=
		\{\rho\in\rhoGrid:E(\rho)\leq\eta\},
		\label{eq:rho-feasible-set}
	\end{equation}
	where \(\eta>0\) is the allowed energy-deviation tolerance.
	
	In the second stage, we further introduce a residual-whiteness criterion within \(\mathcal{C}_{\eta}\). For the vectorized residual, we use the one-dimensional normalized sample autocorrelation
	\begin{equation}
		c_{\ell}\left(\mathbf{r}\right)
		=
		\frac{\sum_{i=1}^{m-\ell} r_{i+\ell}\overline{r_i}}
		{\sum_{i=1}^{m}|r_i|^2},
		\qquad \ell=1,\ldots,L_c,
		\label{eq:residual-autocorr}
	\end{equation}
	where \(L_c\) is the maximum lag used to truncate the test of low-order correlations. Based on these low-order correlation coefficients, we define the whiteness score as
	\begin{equation}
		W(\mathbf{r})
		=
		\sum_{\ell=1}^{L_c}|c_{\ell}(\mathbf{r})|^2.
		\label{eq:whiteness}
	\end{equation}
	When the residual is closer to white noise, its nonzero-lag autocorrelations are closer to zero, and \(W(\mathbf{r})\) becomes smaller.
	
	Finally, \(\rho_k\) is selected as the parameter that gives the residual with the smallest whiteness score among the candidates satisfying energy consistency:
	\begin{equation}
		\rho_k
		=
		\argmin_{\rho\in\mathcal{C}_{\eta}}
		W\left(\mathbf{r}_{k+1}(\rho)\right).
		\label{eq:hybrid-rho}
	\end{equation}

	This hierarchical rule first uses the discrepancy constraint to exclude candidates that clearly disagree with the noise power, and then uses the whiteness criterion to suppress structured correlated components in the residual. It is therefore more robust than using either residual energy or whiteness alone. In implementation, \(\rho_k\) can be found by searching a prescribed candidate set, or by using a one-dimensional automatic search method such as golden-section search.
	
	After the data-consistency update, the prior update replaces the proximal mapping in \eqref{eq:x-prox} with CM denoising. Treating the CM denoiser as an approximate proximal operator suggests a Gaussian denoising scale of \(\sqrt{\lambda/\rho_k}\). The proposed algorithm therefore sets the CM denoising level from the observed \(\snrdb\) and the selected ADMM penalty parameter. Specifically, the regularization strength is modeled as
	\begin{equation}
		\lambda(\snrdb)
		=
		\alpha_{\lambda}10^{-\beta_{\lambda}\snrdb/10},
		\label{eq:lambda-policy}
	\end{equation}
	where \(\alpha_{\lambda}\) and \(\beta_{\lambda}\) control the variation of prior strength with \(\snrdb\). The CM denoising level is then set as
	\begin{equation}
		t_k=\sqrt{\frac{\lambda(\snrdb)}{\rho_k}}.
		\label{eq:tau-rule}
	\end{equation}
	Intuitively, a lower \(\snrdb\) corresponds to a larger \(\lambda(\snrdb)\), thereby strengthening the learned prior; a higher \(\snrdb\) corresponds to a smaller regularization strength, which helps avoid over-smoothing. In practice, upper and lower bounds can also be imposed on \(\lambda(\snrdb)\) to prevent the prior strength from becoming too large or too small under extreme \(\snrdb\) conditions.
	
	After the prior update, the algorithm uses a fixed momentum extrapolation to accelerate convergence:
	\begin{align}
		\widehat{\mathbf{x}}_{k+1}
		&=
		\mathbf{x}_{k+1}
		+
		\beta_m(\mathbf{x}_{k+1}-\mathbf{x}_k),\\
		\widehat{\boldsymbol{\mu}}_{k+1}
		&=
		\boldsymbol{\mu}_{k+1}
		+
		\beta_m(\boldsymbol{\mu}_{k+1}-\boldsymbol{\mu}_k).
	\end{align}

	The overall algorithm is summarized in Algorithm~\ref{alg:adaptive-pnp-cm}.
	
	\begin{algorithm}[t]
		\caption{Proposed Channel Estimation Algorithm}
		\label{alg:adaptive-pnp-cm}
		\begin{algorithmic}[1]
			\REQUIRE \(\mathbf{y}\), \(\mathbf{A}\), \(\sigma_n^2\), \(\snrdb\), CM denoiser \(f_{\theta}\), grid \(\rhoGrid\), \(K\), \(\eta\), \(L_c\), \(\alpha_{\lambda}\), \(\beta_{\lambda}\), \(\beta_m\)
			\STATE Initialize \(\mathbf{x}_0=\boldsymbol{\mu}_0=\widehat{\mathbf{x}}_0=\widehat{\boldsymbol{\mu}}_0=\mathbf{0}\)
			\STATE \(\lambda=\alpha_{\lambda}10^{-\beta_{\lambda}\snrdb/10}\)   
			\FOR{\(k=0,\ldots,K-1\)}
			\STATE Obtain \(\rho_k\) from \eqref{eq:hybrid-rho}
			\STATE \(\mathbf{z}_{k+1}=(\mathbf{A}^H\mathbf{A}+\rho_k\Id)^{-1}(\mathbf{A}^H\mathbf{y}+\rho_k(\widehat{\mathbf{x}}_k-\widehat{\boldsymbol{\mu}}_k))\)
			\STATE \(t_k=\sqrt{\lambda/\rho_k}\)
			\STATE \(\mathbf{x}_{k+1}=f_{\theta}(\mathbf{z}_{k+1}+\widehat{\boldsymbol{\mu}}_k,t_k)\)
			\STATE \(\boldsymbol{\mu}_{k+1}=\widehat{\boldsymbol{\mu}}_k+\mathbf{z}_{k+1}-\mathbf{x}_{k+1}\)
			\STATE \(\widehat{\mathbf{x}}_{k+1}=\mathbf{x}_{k+1}+\beta_m(\mathbf{x}_{k+1}-\mathbf{x}_k)\)
			\STATE \(\widehat{\boldsymbol{\mu}}_{k+1}=\boldsymbol{\mu}_{k+1}+\beta_m(\boldsymbol{\mu}_{k+1}-\boldsymbol{\mu}_k)\)
			\ENDFOR
			\ENSURE \(\widehat{\mathbf{h}}=\mathbf{x}_K\)
		\end{algorithmic}
	\end{algorithm}

	\subsection{CM Network Architecture}
	\label{subsec:cm-network}
	
\begin{figure}[t]
		\centering
		\includegraphics[width=0.98\columnwidth]{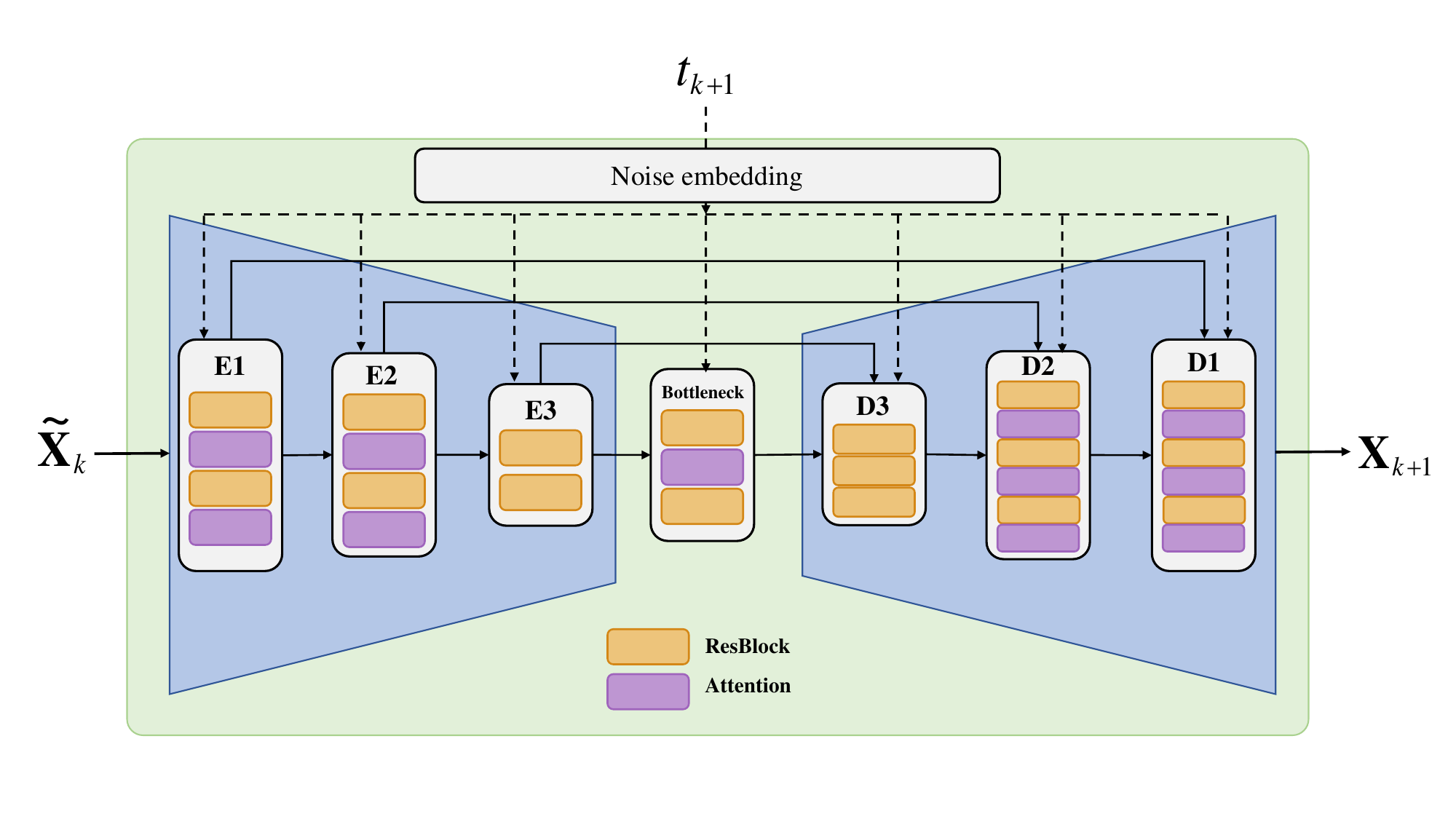}
		\caption{Conditional U-Net denoiser used by the CM prior.}
		\label{fig:cm-architecture}
	\end{figure}

	The denoiser \(f_{\theta}\) is implemented as the lightweight conditional U-Net shown in Fig.~\ref{fig:cm-architecture}. The network input \(\widetilde{\mathbf{X}}_{k}\) is obtained by reshaping the complex channel estimate \(\widetilde{\mathbf{x}}_{k}=\mathbf{z}_{k+1}+\widehat{\boldsymbol{\mu}}_k\) into the angle-domain matrix form and stacking its real and imaginary parts as two channels. This two-channel tensor, together with the denoising level \(t_k\), is fed into the conditional U-Net.
	
	The encoder progressively reduces the spatial resolution and extracts higher-level angular-domain features. Each level consists of several residual blocks with condition injection, and self-attention modules are inserted at selected scales to capture correlations between distant angular cells. The decoder progressively upsamples the feature maps and concatenates the corresponding encoder features through skip connections to recover fine-grained angular-domain structure. The final convolution maps the features back to the two-channel real/imaginary output. The CM denoising level is explicitly fed into the network as a condition variable, processed, and injected into each residual block. The implemented \(f_{\theta}\) wraps the U-Net backbone \(F_{\theta}\) with a CM preconditioning layer:
	\begin{equation}
		f_{\theta}(\mathbf{X},t_k)
		=
		c_{\mathrm{skip}}(t_k)\mathbf{X}
		+
		c_{\mathrm{out}}(t_k)
		F_{\theta}\!\left(c_{\mathrm{in}}(t_k)\mathbf{X},c_{\mathrm{noise}}(t_k)\right).
	\end{equation}
	Here, \(c_{\mathrm{skip}}\), \(c_{\mathrm{out}}\), and \(c_{\mathrm{in}}\) are determined by the training-data standard deviation and the minimum-noise boundary, and satisfy \(c_{\mathrm{skip}}(\epsilon)=1\) and \(c_{\mathrm{out}}(\epsilon)=0\) \cite{song2023consistency,song2023improved}.

	\section{Simulation Results}
	\label{sec:experiments}
	
	\subsection{Datasets and Parameter Settings}
	
	The network training and main experiments use a \(60\) GHz narrowband MIMO angle-domain channel dataset generated by QuaDRiGa under the mmMAGIC\_UMi\_LOS scenario \cite{jaeckel2014quadriga}. The training, validation, and test sets contain \(8000\), \(1000\), and \(1000\) samples, respectively. Cross-dataset generalization is evaluated on \(1800\) test samples generated from the S002 subset of Raymobtime \cite{klautau20185g}.
	
	Each channel sample has \(N_r=16\) receive antennas, \(N_t=64\) transmit antennas, and \(L_r=2\) radio-frequency (RF)-chain outputs. The phases of the transmit pilot matrix \(\mathbf{P}\) and the receive combining matrix \(\mathbf{W}\) are quantized by \(4\)-bit finite-resolution phase shifters. The pilot ratio is \(\alpha=M/(N_tN_r)=M_tM_rL_r/(N_tN_r)\). Estimation accuracy is measured by the normalized mean squared error (NMSE)
	\begin{equation}
		\nmse_{\mathrm{dB}}
		=
		10\log_{10}
		\frac{\|\widehat{\mathbf{h}}-\mathbf{h}\|_2^2}
		{\|\mathbf{h}\|_2^2}.
	\end{equation}
	
	Unless otherwise specified, inference uses \(K=10\) outer PnP iterations. The \(\rho_k\) search grid contains \(40\) logarithmically spaced candidates in \([0.002,30]\), with \(\eta=0.3\), \(L_c=8\), \(\alpha_{\lambda}=1.05\), \(\beta_{\lambda}=0.8\), and \(\beta_m=0.06\). For training the CM, the Karras exponent is \(7\), the discretization increases from \(s_0=10\) to \(s_1=640\), \(P_{\mathrm{mean}}=-0.3466\), \(P_{\mathrm{std}}=1.2\), the pseudo-Huber coefficient is \(0.774\), and \(\epsilon=0.05\), \(\sigma_{\max}=3.2\), \(\sigma_{\min}=0.001\). Other training details follow \cite{song2023consistency,song2023improved}.
	
	\subsection{Performance Evaluation}
	
\begin{figure}[t]
		\centering
		\includegraphics[width=0.98\columnwidth]{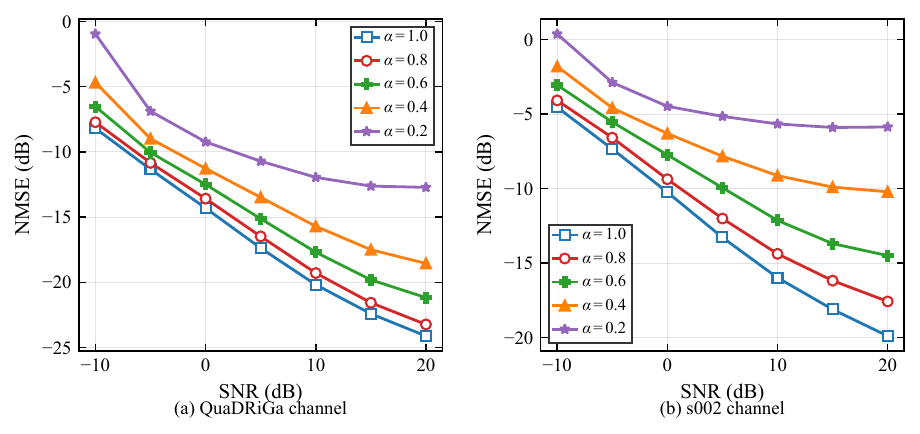}
		\caption{NMSE versus SNR for different pilot ratios on the QuaDRiGa and S002 datasets. The number of outer PnP iterations is \(K=10\).}
		\label{fig:main-performance}
	\end{figure}

	Fig.~\ref{fig:main-performance}(a) shows that the NMSE of the proposed algorithm decreases steadily as SNR increases on the in-distribution QuaDRiGa test set; at pilot ratio \(0.8\), its NMSE remains close to the full-pilot case, while lower pilot ratios suffer larger high-SNR degradation due to the unobserved subspace. Fig.~\ref{fig:main-performance}(b) applies the QuaDRiGa-trained prior directly to S002 channels. Although the curves shift upward under dataset mismatch, the algorithm still preserves useful performance, indicating robust cross-dataset generalization.
	
\begin{figure}[t]
		\centering
		\includegraphics[width=0.98\columnwidth]{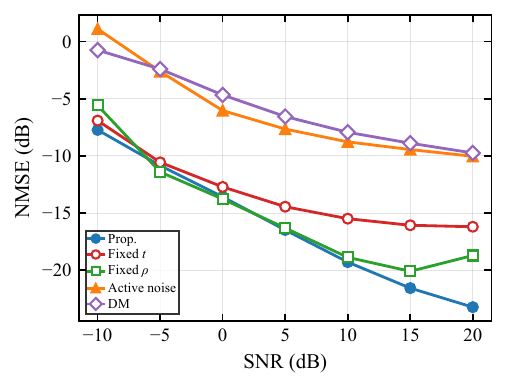}
		\caption{Ablation and baseline comparison on QuaDRiGa at pilot ratio \(0.8\). The number of outer PnP iterations is \(K=10\).}
		\label{fig:ablation}
	\end{figure}

	Fig.~\ref{fig:ablation} compares the proposed algorithm, two core ablations, an active-noise variant, and a diffusion-model baseline. The fixed-\(t\) ablation keeps adaptive \(\rho_k\) but fixes \(\{t_k\}_{k=1}^{K}\) using a random sample at \(\snrdb=0\) dB and pilot ratio \(0.6\). The fixed-\(\rho\) ablation fixes \(\{\rho_k\}_{k=1}^{K}\) from a random sample, while retaining the \(\snrdb\)-scaled \(\lambda\) and \(t_k=\sqrt{\lambda/\rho_k}\). These settings isolate adaptive CM denoising level and adaptive data-consistency weighting.
	
	The proposed algorithm is the most stable across SNRs and has a clear high-SNR advantage. The fixed-\(t\) curve saturates after \(10\) dB, showing that a static CM denoising level over-smooths details when pilot observations become reliable. The fixed-\(\rho\) curve is close to the proposed algorithm around \(-5\) and \(0\) dB, but degrades after \(15\) dB, indicating that one reference penalty sequence cannot match all noise levels.
	
	The active-noise variant keeps the proposed rules for \(\rho_k\) and \(t_k\), but injects complex Gaussian noise into \(\mathbf{z}_{k+1}+\widehat{\boldsymbol{\mu}}_k\) before each CM update; the injected standard deviation and CM denoising level are both set to \(3t_k\). The diffusion baseline uses the prior in \cite{fesl2024diffusion}: it first computes \(\mathbf{z}\) through the proposed data-consistency step and then feeds \(\mathbf{z}\) into the diffusion denoiser.
	
	The active-noise and diffusion variants are significantly weaker than the proposed algorithm. Active noise injection is sensitive because the injected standard deviation must match the SNR, pilot ratio, observation conditioning, and CM operating range. The diffusion baseline is also mismatched to this compressed pilot model: its denoiser is designed for full-pilot or approximately additive white Gaussian noise (AWGN)-perturbed least-squares (LS) observations. Here, even the regularized data-consistency output \(\mathbf{z}\) contains colored, biased, and partially unobservable errors, and a single diffusion denoising step has no subsequent measurement-consistency projection.

	\begin{figure}[t]
		\centering
		\includegraphics[width=0.98\columnwidth]{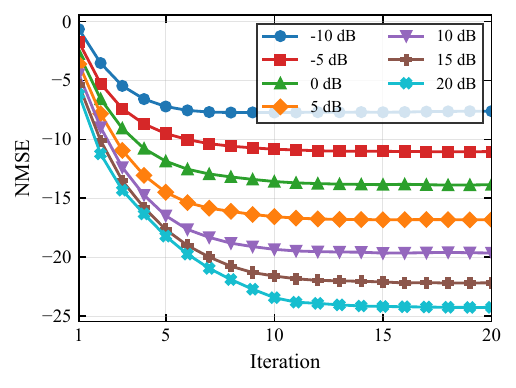}
		\caption{Iteration-wise NMSE of the proposed algorithm on QuaDRiGa at pilot ratio \(0.8\).}
		\label{fig:iteration}
	\end{figure}
	
	Fig.~\ref{fig:iteration} reports iteration-wise NMSE for the proposed algorithm. Most gains appear within the first few iterations. For medium \(\snrdb\) values, \(5\) to \(10\) iterations are sufficient; low-\(\snrdb\) curves saturate earlier because observation noise dominates, whereas high-\(\snrdb\) curves still benefit from additional refinement. Thus, \(K=10\), used in Figs.~\ref{fig:main-performance} and \ref{fig:ablation}, balances accuracy and inference cost. This iteration count is also below the 20 to 100 ODE reverse steps reported by DiffPace \cite{hu2026diffpace}. Since the current experiments use a fixed momentum coefficient for all SNRs and iterations, adaptive momentum may further reduce the required number of iterations without sacrificing final NMSE.

	\section{Conclusion}
	\label{sec:conclusion}
	
	In this paper, we investigated CM-based channel estimation for MIMO systems under compressed pilot observations. First, we designed and trained a conditional U-Net-based CM to learn the angle-domain channel distribution, and embedded the learned generative prior into the PnP-ADMM framework. We also proposed an adaptive parameter-selection mechanism that determined the ADMM penalty parameter from residual energy and residual whiteness, and set the CM denoising level according to the observed SNR and the selected penalty parameter. Simulation results showed that the proposed algorithm could achieve high estimation accuracy with substantially fewer inference steps, and maintained favorable performance across different pilot ratios and under cross-dataset generalization. Future work will further explore adaptive momentum and broader channel conditions to improve convergence speed and robustness.

	\bibliographystyle{IEEEtran}
	\bibliography{references}
	
\end{document}